\documentclass[sigconf,screen,authorversion,nonacm]{acmart}

\AtBeginDocument{%
  \providecommand\BibTeX{{%
    \normalfont B\kern-0.5em{\scshape i\kern-0.25em b}\kern-0.8em\TeX}}}



\makeatletter
\def\@ACM@checkaffil{
    \if@ACM@instpresent\else
    \ClassWarningNoLine{\@classname}{No institution present for an affiliation}%
    \fi
    \if@ACM@citypresent\else
    \ClassWarningNoLine{\@classname}{No city present for an affiliation}%
    \fi
    \if@ACM@countrypresent\else
        \ClassWarningNoLine{\@classname}{No country present for an affiliation}%
    \fi
}
\makeatother

\usepackage{tikz}
\usepackage{subcaption}
\usepackage{listings}
\usepackage{circledsteps}
\usepackage{numprint}
\usepackage{xspace}
\usepackage{multirow}
\usepackage{threeparttable}
\usepackage{soul}
\usepackage{xargs}
\usepackage{enumitem}
\usepackage{longtable}
\usepackage{array}
\usepackage{booktabs}
\usepackage{listing}
\usepackage{xspace}
\usepackage{xargs}
\usepackage{xstring}
\usepackage{hyperref}
\usepackage{textcomp}
\usepackage{xcolor}
\usepackage{lmodern}
\usepackage{graphicx}
\usepackage{enumitem}
\usepackage{amsmath}
\usepackage{mathtools}
\usepackage{multirow}
\newif\ifcomments
\commentstrue
\ifcomments
\newcommand{\todo}[1]{\textcolor{red}{TODO: #1}}
\newcommand{\new}[1]{\textcolor{red}{#1}}
\newcommand{\Chao}[1]{\textcolor{red}{Chao: #1}}
\newcommand{\Zhun}[1]{\textcolor{red}{Zhun: #1}}
\newcommand{\Weilin}[1]{\textcolor{red}{Weilin: #1}}
\newcommand{\Dayu}[1]{\textcolor{red}{Dayu: #1}}
\else
\newcommand{\todo}[1]{}
\newcommand{\new}[1]{}
\newcommand{\Chao}[1]{}
\newcommand{\Zhun}[1]{}
\newcommand{\Weilin}[1]{}
\newcommand{\Dayu}[1]{}

\fi

\begin{document}

\title{Unmasking Role-Play Attack Strategies in Exploiting Decentralized Finance (DeFi) Systems}
\titlenote{This work is supported in part by the National Key Research and Development Program of China (2021YFB2701000).}
\author{Weilin Li}
\affiliation{\small{University of Science and Technology of China}}
\author{Zhun Wang}
\affiliation{\small{Tsinghua University}}
\author{Chenyu Li}
\affiliation{\small{Institute of Information Engineering Chinese Academy of Sciences}}
\author{Heying Chen}
\affiliation{\small{University of Science and Technology of China}}
\author{Taiyu Wong}
\affiliation{\small{Tsinghua University}}
\author{Pengyu Sun}
\affiliation{\small{University of Science and Technology of China}}
\author{Yufei Yu}
\affiliation{\small{Tsinghua University}}
\author{Chao Zhang}
\affiliation{\small{Tsinghua University}}


\renewcommand{\shortauthors}{Li et al.}


\begin{abstract}
\vspace{-0.1cm}
The rapid growth and adoption of decentralized finance (DeFi) systems have been accompanied by various threats, notably those emerging from vulnerabilities in their intricate design. In our work, we introduce and define an attack strategy termed as \textit{Role-Play Attack}, in which the attacker acts as multiple roles concurrently to exploit the DeFi system and cause substantial financial losses. We provide a formal definition of this strategy and demonstrate its potential impacts by revealing the total loss of \$435.1M caused by 14 historical attacks with applying this pattern. Besides, we mathematically analyzed the attacks with top 2 losses and retrofitted the corresponding attack pattern by concrete execution, indicating that this strategy could increase the potential profit for original attacks by \textbf{\$3.34M (51.4\%)} and \textbf{\$3.76M (12.0\%)}, respectively.
\end{abstract}


\maketitle

\vspace{-0.26cm}
\section{Introduction}\label{sec:Introduction}

Blockchain technology has deeply impacted the financial technology environment~\cite{makridis2023rise}. Its revolutionary potential led to the development of decentralized finance (DeFi), a promising financial system~\cite{qin2021cefi}. Nevertheless, the unique complexity of DeFi systems, like any emerging technology, entails an array of potential vulnerabilities that could be exploited~\cite{werner2022sok}. These flaws primarily compromise implementation flaws resulting from coding issues and logic flaws resulting from poor designs. Rectifying implementation vulnerabilities usually involves code-focused analysis, while addressing logic vulnerabilities often requires mathematical and financial expertise.

Against this intricate backdrop of potential threats, we delve into various historical attacks instigated by logic vulnerabilities. A careful study of these incidents led us to identify a recurrent pattern, which we have termed \textit{Role-Play Attack}. This attack strategy involves an attacker acting as various roles (e.g., lender, borrower, trader) simultaneously, exploiting the DeFi system to achieve substantial financial benefits. Instead of aiming at certain vulnerabilities like re-entrancy attacks, \textit{Role-Play Attack } emphasizes the combination of roles and actions involved in an attack.
In light of this, we present a formal definition of \textit{Role-Play Attack } in Section \ref{sec:define} and delve into two historical attack methods in Section \ref{sec:history}. 
In Section \ref{sec:enhance}, we prove that the attacker's profits could be further promoted with this strategy and analyze the possible maximum gains. We propose our conclusion in Section \ref{sec:conclusion} and point out two potential future works to further analyze or mitigate this attack in 
 Section \ref{sec:discussion}.

The main contributions of this paper are summarized as follows:
\vspace{-0.25cm}
\begin{itemize}[leftmargin=*]
\item \textbf{Formal Definition of \textit{Role-Play Attack}:} We propose a definition for a recurrent attack strategy, the \textit{Role-Play Attack}. Through our data collection and manual inspection, we identified this strategy in 14 distinct DeFi security incidents, leading to an accumulated financial loss of  \$435.1 million\footnote{This accounts for around 6.38\% compared with the total hacked value (THV) of \$ 6.82B according to \hyperlink{https://defillama.com/hacks}{https://defillama.com/hacks}.}.
\item \textbf{Comprehensive Analysis of \textit{Role-Play Attack}:} We conduct a detailed exploration of the role-play strategy by analyzing two distinct attack methods with mathematical analysis. Our analysis brings forth the roles and their specific impacts, unveiling the complex dynamics underlying such malicious exploits.
\item \textbf{Promoting Profits of \textit{Role-Play Attack }:} We delve into two historical attack events' 
profitability from an attacker's viewpoint. Using advanced modeling and analysis, we have achieved a significant improvement in the attacker's profits from two historical attacks, by \textbf{\$3.34M (51.4\%)} and \textbf{\$3.76M (12.0\%)}, respectively.
\end{itemize}

\vspace{-0.3cm}
\section{Background and Related Works}\label{sec:Background}
The inception of blockchain technology, distinguished by its decentralization and distributed ledger capabilities, brought forth a new paradigm for transparent and tamper-resistant transaction recording~\cite{nakamoto2009bitcoin}. Blockchain, which is essentially a permissionless peer-to-peer (P2P) network, uses Proof of Work (PoW)~\cite{gervais2016security, biryukov2017proofs} and Proof of Stake (PoS)~\cite{grandjean2023ethereum} as separate consensus procedures and allows any member to transmit transactions~\cite{wust2018you}. Notably, the emergence of smart contracts extended the financial utility of blockchain, catalyzing the rapid growth of DeFi with a peak historic total value locked (TVL) exceeding \$150 billion~\cite{werner2022sok}.
\vspace{-0.25cm}
\subsection{Decentralized Finance (DeFi)}
Decentralized finance (DeFi) is a novel, permissionless financial system employing blockchain technology to execute operations in a transparent, decentralized manner~\cite{antonopoulos2014mastering, consensys2021}. At its core, DeFi relies on autonomous smart contracts to maintain transaction transparency and immutability~\cite{antonopoulos2014mastering}. By facilitating peer-to-peer transactions, DeFi aims to establish robust alternatives to traditional finance~\cite{consensys2021, qin2021cefi}. Readers seeking a comprehensive understanding of DeFi are referred to existing literature, notably  reference~\cite{bartoletti2021towards, werner2022sok}. Three of the primary infrastructures within the DeFi ecosystem are as follows:
\vspace{-0.1cm}
\begin{itemize}[leftmargin=*]
\item \textbf{Lending:} In DeFi lending markets, users have the chance to lend their assets to earn interest, resembling traditional finance mechanisms where debt is a crucial instrument~\cite{bartoletti2021sok}. The interest rates, typically determined by supply-borrow dynamics~\cite{xu2022banks}, generate a revenue stream. 
To safeguard lenders, borrowers are often required to provide more collateral than the loan value, which can be liquidated if its value dips~\cite{qin2021empirical}.
\item \textbf{Decentralized Exchange (DEX):} DEXes allow users to trade their assets for another through liquidity pools. Instead of using the classic order book model where buyers and sellers place orders at their preferred prices, DEXes often use Auto Market Makers (AMMs)~\cite{xu2023sok}. AMMs algorithmically set the price of tokens, offering a seamless and efficient trading experience that doesn't require matching individual buy and sell orders.
\item \textbf{Yield:} DeFi yielding protocols, often built on top of lending markets or DEXes, offer users with interests to earn from farming their assets. Profits often come from interest earned and rewards in the form of additional tokens. To optimize the profits, different yield farming strategies have been adopted in practical use~\cite{cousaert2022sok}.
\end{itemize}
\vspace{-0.25cm}
\subsection{DeFi Attack}
The exponential growth of DeFi has made it an alluring target for hackers~\cite{gervais2023security}. Between April 30, 2018, and April 30, 2022, DeFi protocols experienced losses exceeding \$3 billion due to various attacks~\cite{zhou2022sok}. Noteworthy attack types include re-entrancy attacks, flash-loan attacks, and oracle manipulation attacks, with each exploiting different facets of the DeFi system vulnerabilities. 
Numerous studies have addressed various forms of attacks~\cite{eskandari2020sok, zhou2021high, li2022security, yaish2023speculative}. Remarkably, Qin et al. systematically analyzed flash loan attacks and performed optimizations on historical attack incidents~\cite{qin2021attacking}.
\vspace{-0.25cm}
\subsection{Security of DeFi}
The security of DeFi is multifaceted, addressing both code-related vulnerabilities and economic design flaws.
From the perspective of smart contract code vulnerabilities, several detection tools have been developed. These tools utilize static and dynamic analysis to detect potential issues~\cite{chaliasos2023smart}. Static analysis examines the contract's code without executing it, aiming to find vulnerabilities through code patterns and flow analysis~\cite{tsankov2018securify, sharmasurvey}. On the other hand, dynamic analysis involves executing the contract's code in a controlled environment to monitor the runtime behavior~\cite{rodler2018sereum, mossberg2019manticore, chen2020soda, qin2023auto}. 

The realm of DeFi security extends beyond mere code vulnerabilities~\cite{qin2021cefi, werner2022sok}. The design of the underlying economic mechanisms can also introduce exploitable weak points~\cite{Cohen23liq}. It is important to take a comprehensive approach to DeFi security, considering both code-related and economic risks.

\vspace{-0.2cm}
\section{Definitions and Models}\label{sec:define}
\subsection{Formal Definition}
As we delve into the exploration of the Role-Play Attack, it is paramount to begin by setting a clear understanding of the terminologies and concepts that form the foundation of such an attack strategy.
\vspace{-0.2cm}
\subsubsection{System Models} This section presents the conceptual framework for understanding \textit{Role-Play Attacks}.

\vspace{-0.1cm}
\begin{itemize}[leftmargin=*]
    \item \textbf{Smart Contract Set}: A smart contract set is an extensive collection of linked smart contracts, indicated as: 
    \vspace{-0.1cm}
    \[S = \{s_1, s_2, \dots\}\]
    Each contract $s_i$ within this set forms an integral part of the structure and is interconnected with other contracts through various relationships like function calls and data dependency.
    \item \textbf{Function Call Sequence} ($c$): Given a set of external functions:
    \vspace{-0.1cm}
    \[F = \bigcup_{i}\ \{f | f\ \text{is an external function of}\ s_i, s_i \in S\}\] 
    associated with the contracts of a protocol, a function call sequence $c$ is defined as an ordered sequence of $F$. The collection of all possible function call sequences is denoted as: 
    \[C(F) = \bigcup_{i \in \mathbb{Z}_{>0}} F^i, \ c = (f_1, f_2, \dots) \in C(F), \ f_j \in F\]

    \item \textbf{Actions} ($A$) \textbf{and Events} ($E$): An action $A$ is a function call sequence that achieves a purposeful activity. A common approach involves utilizing the \textit{approve} and \textit{borrow} functions to execute a borrowing action. An event $E$ comprises a sequence of actions:
    \vspace{-0.15cm}
    \[E = (A_1, A_2, \dots)\]
    \item \textbf{Gains} ($g$): The cumulative gain of an address $a$ involved in an event $E$, denoted as $g_E(a)$, is the sum of the profit $w_{A_i}(a)$ of the address (denominated in USD) from each action $A_i$ within $E$, i.e.,
    \vspace{-0.15cm}
    \[g_E(a) = w_{A_1}(a) + w_{A_2}(a) + \dots\]
    \item \textbf{Roles} ($r$): A role $r$ is an address that performs particular actions. Examples of roles include borrowers, traders and lenders. 
\end{itemize}

\vspace{-0.2cm}
\subsubsection{Threat Models}

\begin{center}
    \begin{figure}[t]
  \raggedright
  \includegraphics[width=0.48 \textwidth]{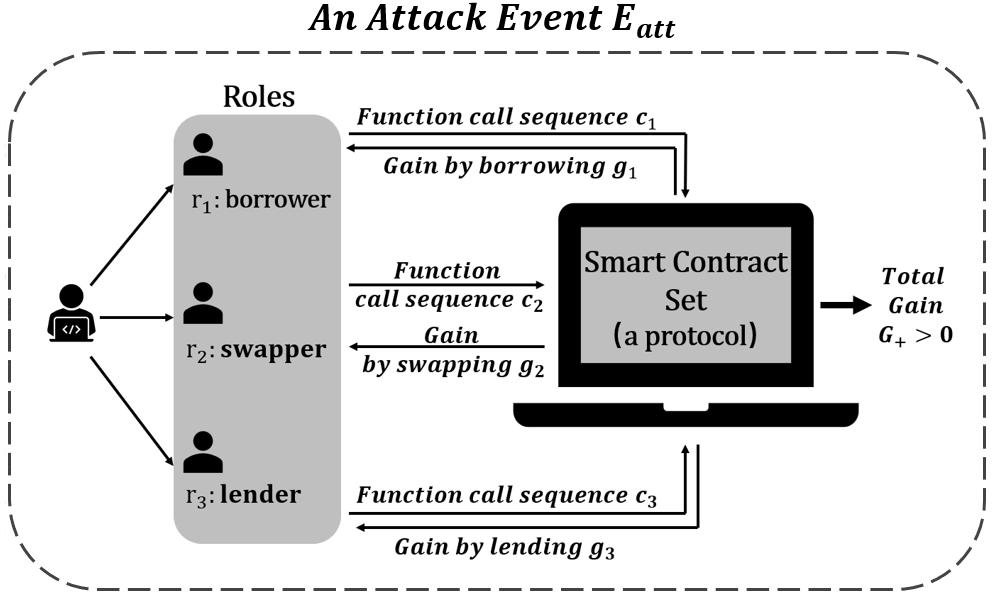} 
  \vspace{-0.6cm}
  \caption{An example of \textit{Role-Play Attack }} 
  \vspace{-0.45cm}
  \label{fig:Eatt} 
\end{figure}
\end{center}

Let $E_{att}$ represent a \textit{Role-Play Attack}, as shown in Figure~\ref{fig:Eatt}, We can define the threat model as the followings.

\begin{itemize}[leftmargin=*]
    \item \textbf{Multiple roles activities}:  The attacker $R$ engages in a malicious activity by acting as multiple roles $r_i$, i.e.,
    \vspace{-0.1cm}
    \[R = \{r_1, r_2, \dots, r_n\}\]
    \item \textbf{The result}: The gain $G_a$ of the attacker is the cumulative  sum of the gains of the roles (some of which can be negative), i.e.,
    \vspace{-0.1cm}
    \[G_a = g_{E_{att}}(r_1) + g_{E_{att}}(r_2) + \dots + g_{E_{att}}(r_n)\]
     A normal user benefits from the DeFi system with gains $G_n$, while an attacker of a \textit{Role-Play Attack } can achieve significantly greater gains with $G_a >> G_n$.
\end{itemize}
%
%
\vspace{-0.6cm}
\subsection{Common Roles in Different DeFi Systems}

DeFi presents a wide array of services to users, enabling a multifaceted engagement within each specific DeFi ecosystem. Depending on the system type, users can adopt various roles. Some notable systems and their corresponding user roles are described below. The functions of each role are shown in Table~\ref{tb:defi}.
\begin{table}[t]
\begin{tabular}{|l|ll|}
\hline
Protocol                 & Role                                                         & Function                                             \\ \hline
\multirow{3}{*}{Lending} & Lender                                                       & {\tt lend(asset, amount)}               \\
                         & Borrower                                                     & {\tt borrow(asset, amount)}             \\
                         & Liquidator                                                   & {\tt liquidate(borrower, debt)} \\ \hline
\multirow{2}{*}{DEX}     & Trader                                                       & {\tt swap(token1, token2, amount)}   \\
                         & \begin{tabular}[c]{@{}l@{}}Liquidity\\ Provider\end{tabular} & {\tt addLiquidity(pool, amount)}        \\ \hline
\multirow{2}{*}{Yield}   & Yield Farmer                                                 & {\tt claimReward()}                     \\
                         & Yield Source                                                 & {\tt addReward(amount)}                 \\ \hline
\end{tabular}
\caption{Different roles in common DeFi protocols.}\label{tb:defi}
\vspace{-0.8cm}
\end{table}

\vspace{-0.15cm}
\begin{itemize}[leftmargin=*]
    \item \textbf{Lending Markets}
\begin{itemize}[leftmargin=*, label=]
    \item \textbf{Lender:} A user who deposits assets into the market to be loaned, typically receiving interest payments as a reward.
    \item \textbf{Borrower:} A user who obtains assets from the market, ordinarily providing collateral to safeguard the loan.
    \item \textbf{Liquidator:} A user responsible for reimbursing a borrower's debt when the value of the collateral falls short, thereby maintaining the system's balance and stability.
\end{itemize}
    
    \item \textbf{DEX}
    \begin{itemize}[leftmargin=*, label=]
        \item \textbf{Trader:} A user who engages in exchanging assets.
        \item \textbf{Liquidity Provider:} A user who contributes funds to a liquidity pool. These funds are then used to facilitate trading activities within the liquidity pool, earning swap fees in return.
    \end{itemize}
    
    \item \textbf{Yield Farming}
    \begin{itemize}[leftmargin=*, label=]
        \item \textbf{Yield Farmer:} A user who commits assets to a DeFi protocol with the intention of earning rewards.
        \item \textbf{Yield Source:} Typically a DeFi protocol offering rewards to users for depositing or locking their assets. These incentives can emanate from various sources, such as transaction fees.
    \end{itemize}
\end{itemize}
\vspace{-0.2cm}
\section{\textit{Role-Play Attacks}: A Deep Dive}\label{sec:history}
\subsection{\textit{Role-Play Attacks} in history}
\textit{Role-Play Attacks} have been a common tactic among malicious entities intending to exploit DeFi systems, thereby inflicting significant financial losses. To understand the scale of these breaches, we have constructed a dataset comprising 14 \textit{Role-Play Attack} incidents that occurred between September 28, 2020, and May 13, 2023. These attacks caused losses exceeding \$400 million in total, which, as illustrated in Table \ref{tb:attack}, may serve as a conservative estimate of the comprehensive financial loss induced by such attack patterns. The dataset is primarily derived from sources such as \textit{BlockSec}~\cite{blocksec}, \textit{Rekt News Leaderboard}~\cite{rektLeader}, \textit{PeckShield}~\cite{peckshield}, and \textit{SlowMist}~\cite{slowmist}. 

To gain a better understanding of such attack patterns, we delve into this prevalent strategy by examining two case studies in Sections \ref{subsec:bb} and \ref{subsec:bd}. These incidents, whose losses rank top 2 in our dataset, represent two typical types of \textit{Role-Play Attacks}\footnote{$\dagger$: B\&B Attack in Section~\ref{subsec:bb}, $\star$: B\&D Attack in Section~\ref{subsec:bd}.}.
\begin{table}[b]
\vspace{-0.5cm}
\begin{tabular}{cccc}
\toprule
Victim Protocol             & Date        & Losses (USD) & Roles \\
\midrule
Cream Finance $\star$ \S\ref{subsec:bd}  & Oct-27-2021 & 130.0M &  1, 2, 5     \\
Mango markets $\dagger$  \S\ref{subsec:bb}  & Oct-11-2022 & 115.0M &    1, 2, 3   \\
Pancake Bunny     & May-19-2021 & 45.0M  &   3, 4, 5    \\
Vee Finance      & Sep-21-2021 & 34.0M  &    3, 4   \\
Spartan Protocol & May-02-2021 & 30.5M  &   3, 4    \\
Cream Finance (2)   & Aug-30-2021 & 18.8M  &    1, 2, 6   \\
Inverse Finance  & Apr-02-2022 & 15.6M  &   1, 2, 3, 4    \\
Eminence & Sep-28-2020 & 15.0M  &    1, 2, 3   \\
Yearn Finance        & Apr-13-2023 & 11.0M  &   1, 2, 4    \\
Moola Market $\dagger$     & Oct-19-2022 & 8.4M   &    1, 2, 3   \\
Lodestar Finance$\star$ & Dec-10-2022 & 6.5M   &    1, 2, 5   \\
0VIX Protocol $\star$           & Apr-28-2023 & 4.3M   &   1, 2, 5    \\
Autoshark        & May-24-2021 & 745.0K &   3, 4, 5   \\
Sell Token DEX          & May-13-2023 & 250.8K &    1, 2, 3  \\
\midrule
Total & - & 435.1M & -\\
\bottomrule
\end{tabular}
\includegraphics[width=0.47\textwidth]{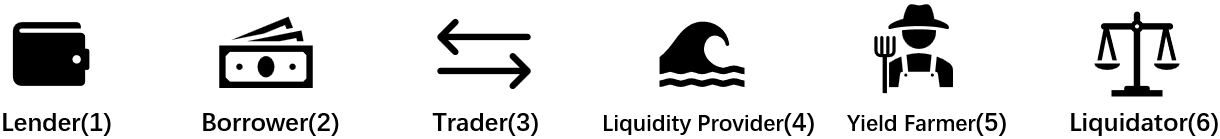}
\vspace{-0.05cm}
\caption{The 14 previous attack incidents with the \textit{Role-Play Attack} pattern. The total losses are worth 435.1M USD.}
\vspace{-0.55cm}
\label{tb:attack}
\end{table}
\vspace{-0.25cm}
\subsection{\textit{Borrow and Buy Attack (B\&B Attack)} \\ - \textit{Role-Play Attack } in Lending and DEX}\label{subsec:bb}
Two salient examples of the \textit{Role-Play Attack } pattern can be found in the Mango Market exploit~\cite{mangoattack} and the Moola Market exploit~\cite{moolaattack}, which rank 17th and 75th on the \textit{Rekt News} \href{https://rekt.news/leaderboard/}{Leaderboard} with losses of \$115 million and \$8.4 million, respectively. These incidents consist of iterations of a common malicious process wherein the attacker repeatedly borrows and purchases crypto assets, a strategy we refer to as the \textbf{Borrow and Buy Attack (B\&B Attack)}.

A further exemplar of this attack pattern is the Agora Lending exploit. In this case, the hacker (identified by the address \href{https://andromeda-explorer.metis.io/address/0xFFD90C77eaBa8c9F24580a2E0088C0C940ac9C48}{0xFFD90C77e-aBa8c9F24580a2E0088C0C940ac9C48}) executed a \textit{B\&B Attack} on Agora Lending~\cite{AgoraFinance}. Throughout this attack, the exploiter fulfilled three roles: \textit{trader}, \textit{lender}, and \textit{borrower}.

\begin{center}
    \begin{figure}[b]
  \raggedright
  \includegraphics[width=0.48 \textwidth]{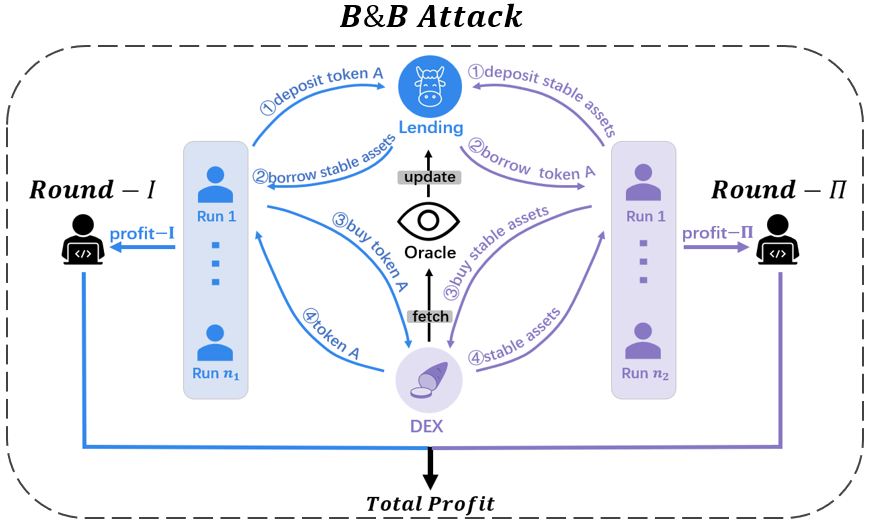} 
  \vspace{-0.4cm}
  \caption{The attack process of a \textit{B\&B Attack}. The left part of the figure shows the initial attack (analyzed in section \ref{subsec:bb}) while the right part illustrades the \textit{Round - II} of the refined attack (analyzed in section \ref{en:bb}).} 
   \label{fig:BB} 
\end{figure}
\end{center}

The Agora Lending attack was halted by the project developers, who lowered the collateral factor of the impacted token to zero~\cite{AgoraHack}. In this section and Section \ref{sec:enhance}, we reconstruct the attack and also explore ways to potentially enhance the attacker's profits by performing reversed attack operations and simulating the roles of a second and subsequent hackers, as shown in Figure \ref{fig:BB}.

\textbf{Overall Attack Process:} To portray the shared pattern of \textit{B\&B Attacks}, we propose a typical attack sequence wherein a lending market allows a low-liquidity token (denoted as token A) to be used as collateral. This attack sequence can be segmented into four steps:
\vspace{-0.15cm}
\begin{enumerate}[leftmargin=*]
\item\label{att:1-A} The attacker bought some token A and deposited it as collateral.
\item\label{att:1-B} The attacker borrowed against this collateral to purchase more of token A, thereby driving up its price. These tokens were subsequently deposited into the lending market.
\item\label{att:1-C} This increase in price and collateral quantity enabled the attacker to borrow more assets.
\item\label{att:1-D} The attacker repeated this process to drain the lending market.
\end{enumerate}
\vspace{-0.1cm}
\textbf{State Assumptions and Models:} To simplify the complexity of real-world DeFi scenarios, we make several assumptions to more effectively analyze the attack process:
\vspace{-0.15cm}
\begin{itemize}[leftmargin=*]
    \item We categorize the funds related to this attack into two types: manipulated and stable assets. The latter refers to funds in the lending pools whose prices remained stable during the attack.
    \item There are three DeFi components involved in this attack: the lending market, the Automated Market Maker (AMM), and the oracle. These components are defined as follows:
    \begin{itemize}[leftmargin=*]
        \item \textit{Lending market}: Facilitates deposits and loans in accordance with borrowing rates and asset prices provided by the oracle.
        \item \textit{AMM}: We consider this AMM to be a Uniswap-V2-style constant product market maker with the equation $xy = constant$~\cite{Adams2020UniswapV2}. The swap fee is disregarded due to its insignificance relative to the potential profits.
        \item \textit{Oracle}: In reality, the oracle extracts the time-weighted average price from the AMM, necessitating the attacker to wait for the oracle's update. For our theoretical analysis, we suppose the attacker waits for a price update after every swap transaction.
    \end{itemize}
    \item For a mathematical analysis of the attack process, we define the following variables:
    \begin{itemize}[leftmargin=*]
        \item $init_{s}$: The initial stable assets in the market (denoted in USD).
        \item $init_{m}$: The initial amount of borrowable manipulated assets in the lending market (denoted in manipulated assets).
        \item $CR_{m}$: The collateral rate of the manipulated assets, which is the ratio between the maximum amount of currency that can be borrowed and the total amount of collateral.
        \item $CR_{s}$: The collateral rate of the stable assets.
        \item $out_{s}$: The required USD value to purchase the manipulated assets in one attack cycle.
        \item $in_{m}$: The amount of purchased assets in one attack cycle.
        \item $L_0$: The value of the stable assets in the pool. Assuming the initial asset price equals one unit of stable assets, the total value of the initial Uniswap V2 liquidity pool is $2L_0$.
    \end{itemize}
\end{itemize}
\vspace{-0.55cm}
\textbf{Mathematical Analysis:} 
First swap some stable assets to manipulated assets, resulting in a new price of manipulated assets:
\vspace{-0.15cm}
\begin{align}
    &p_{new}\ \text{For AMM:}\quad (L_0 - in_{m})\cdot(L_0 + out_{s}) = L_0^2 \notag\\
    &\implies in_{m} = \frac{L_0 \cdot out_{s}}{L_0 + out_{s}} \\
    &\implies p_{\text{new}} = \frac{\text{Reserve}_s}{\text{Reserve}_m} = \frac{L_0 + out_{s}}{L_0 - in_{m}} = \left(1 + \frac{out_{s}}{L_0}\right)^2
\end{align}
The purchased asset can be collateralized to borrow stable assets:
\vspace{-0.15cm}
\begin{align}\label{eq:1-3}
    \text{Borrow Amount} = \overbracket{in_{m} \cdot p_{\text{new}}}^\text{collateral value} \cdot CR_{m} = \left(1 + \frac{out_s}{L_0}\right)\cdot out_s \cdot CR_m
    \end{align}
This attack becomes profitable when:
\begin{itemize}[leftmargin=*]
    \item the value of the borrowed stable assets exceeds the asset purchase amount. (Eq.~\ref{eq:1-4})
    \item there are sufficient stable assets to borrow. (Eq.~\ref{eq:1-5})
\end{itemize}
\vspace{-0.2cm}
\begin{gather}
    \text{Eq.~\ref{eq:1-3}} > out_s \label{eq:1-4}\\
    \text{Eq.~\ref{eq:1-3}} \le init_s \label{eq:1-5}
\end{gather}
When these conditions are met, having Eq.~\ref{eq:1-5} hold as an equality and solving $out_s$ in terms of $CR_m$, $init_s$ and $L_0$, the maximized hacker's profit will be:
\vspace{-0.2cm}
\begin{gather}
    \text{Max Profit} = init_s - out_s = init_s - \left(\sqrt{1 + 4\cdot \frac{init_s}{CR_m\cdot L_0}} - 1\right)\cdot \frac{L_0}{2} \label{eq:1-6}
\end{gather}

\vspace{-0.5cm}
\subsection{\textit{Borrow and Donate Attack (B\&D Attack)}\\ -\ \textit{Role-Play Attack } in Lending and Yield} \label{subsec:bd}
On October 28, 2021, \textit{Cream Finance}~\cite{CreamFinance}, a prominent Ethereum lending platform, was exploited~\cite{creamattack}, resulting in a loss of \$130 million with transaction hash \href{https://etherscan.io/tx/0x0fe2542079644e107cbf13690eb9c2c65963ccb79089ff96bfaf8dced2331c92}{0x0fe2\dots 1c92}. Similarly, \textit{Lodestar Finance}~\cite{LoadstarFinance} fell victim to an exploit~\cite{lodestarattack} on December 12, 2022, costing the platform \$6.5 million as traced in transaction \href{https://arbiscan.io/tx/0xc523c6307b025ebd9aef155ba792d1ba18d5d83f97c7a846f267d3d9a3004e8c}{0xc523\dots 4e8c}. These two attacks rank 13th and 88th on the \textit{Rekt News} \href{https://rekt.news/leaderboard/}{Leaderboard}.

These events shared a unique characteristic: collateral token prices were manipulated via direct transfer of the underlying asset to interest-bearing tokens, a category of tokens that generate yield over time. We refer to this attack as the \textbf{Borrow and Donate Attack (B\&D Attack)}. Here, the attacker played the roles of a \textit{yield farmer}, \textit{lender}, and \textit{borrower}. 

In section \ref{sec:enhance}, we propose enhancements to the \textit{Lodestar Finance} attack operation sequence to reduce flashloan fees and donation amounts, thereby potentially increasing the attacker's profit. We also introduce the \textit{liquidator} role, which can potentially augment the attacker's profit through liquidation rewards. The processes of the primitive attack and the enhanced attack are shown in Figure \ref{fig:BD}.

\textbf{Root Cause:} The \textit{Cream Finance} attack primarily originated from the price manipulation of \textit{Yearn's yUSD vault token} (address: \href{https://etherscan.io/token/0x4B5BfD52124784745c1071dcB244C6688d2533d3}{0x4B5B\dots 33d3}). The attacker repetitively deposited and borrowed the interest-bearing token, effectively inflating the market size to yield significant profits.

\textbf{Overall Attack Process:} In such attacks, the attacker employed two smart contracts, referred to as A and B. Contract A was responsible for minting cryUSD (\textit{Cream Finance}'s lending token for yUSD, address: \href{https://etherscan.io/token/0x4BAa77013ccD6705ab0522853cB0E9d453579Dd4}{0x4BAa\dots 9Dd4}, denoted as $Ctoken_{IB}$) and manipulating Yearn's yUSD vault token (denoted as $token_{IB}$), while Contract B was primarily used to borrow assets against ETH collateral:
\vspace{-0.15cm}
\begin{enumerate}[leftmargin=*]
\item\label{att:2-A} Contract A initiated the attack by borrowing \$500M using a flashloan and depositing the $token_{IB}$ into \textit{Cream Finance}, minting an equivalent amount of $Ctoken_{IB}$.
\item\label{att:2-B} Contract B then borrowed \$2B worth of ETH via flashloan and deposited it into \textit{Cream Finance} as collateral against borrowed $token_{IB}$ from the lending market. This borrowed $token_{IB}$ was then transferred to contrac A and deposited back to mint more $Ctoken_{IB}$. This process was repeated, each time transferring the borrowed $token_{IB}$ to contract A.
\item\label{att:2-C} After repeated mints and borrows, contract A withdrew all the withdrawable $token_{IB}$ from the lending market and burned them to decrease the total supply of $token_{IB}$. A then purchased approximately \$8M worth of $token_{IB}$'s underlying asset to donate and subsequently inflate the price of $token_{IB}$.
\item\label{att:2-D} Following process \ref{att:2-C}, \textit{Cream Finance} was left with a vast quantity of bad debt (around \$3B) on contract B. Contract A then borrowed \$2B worth of assets against the inflated collateral, repaid the initial flash loans to conclude the attack.
\end{enumerate}
\vspace{-0.15cm}
\textbf{State Assumptions and Models:} Similar to section \ref{subsec:bb}, we simplify the model under the following assumptions:
\begin{itemize}[leftmargin=*]
    \item We ignore the friction of swapping the flashloan to the underlying asset of the interest-bearing token.
    \item The initial attack was split into various transactions due to gas limit. In our analysis, we combine them into one transaction.
    \item We define the following variables in our mathematical analysis:
    \begin{itemize}[leftmargin=*]
        \item $supply_{IB}$: The initial total supply of the interest-bearing token.
        \item $borrowable_{IB}$: The initial borrowable interest-bearing token in the lending market.
        \item $borrowable_{s}$: The initial borrowable assets other than the interst-bearing token in the lending market.
        \item $CR_{IB}, CR_{s}$: The collateral rate of the interest-bearing token and stable assets.
    \end{itemize}
\end{itemize}

\textbf{Mathematical Analysis:}
The attack process commences with a flashloan borrow of a significant amount denoted by $flash_{total}$. The fee rate of this flashloan is represented as $flash_{fee}$; hence, the amount to return would be:
\vspace{-0.2cm}
\begin{align}
\begin{aligned}\label{eq:2-1}
flash_{total}/(1-flash_{fee})
\end{aligned}
\end{align}
The flashloan funds are divided into three portions:
\begin{itemize}[leftmargin=*]
\item $init_{mint}$: the funds used to swap for underlying tokens and mint interest-bearing tokens.
\item $collateral_B$: the fund designated as the collateral for Contract B.
\item $donate$: the funds later used for donation purposes.
\end{itemize}

We define $iter$ as the number of iterations. After process \ref{att:2-B}, the collateral of Contract B became $collateral_B$ while Contract B's debt was as given in Eq.~\ref{eq:2-2}. Contract A's collateral is illustrated in Eq.~\ref{eq:2-3} and carries no debt.
\vspace{-0.15cm}
    \begin{align}
    \label{eq:2-2}
    iter \cdot (borrowable_{IB} + init_{mint})\\ 
    init_{mint}+iter\cdot(borrowable_{IB} + init_{mint})\label{eq:2-3}
    \end{align}
The collateral rate of Contract B prior to the donation is shown in Eq.~\ref{eq:2-4}. After donating $donate$ amount of underlying tokens, the price was amplified by a factor of $\epsilon$ (Eq.~\ref{eq:2-5}), thereby causing a substantial amount of bad debt for Contract B.
\vspace{-0.15cm}
    \begin{align}\label{eq:2-4}
    \frac{iter\cdot(borrowable_{IB} + init_{mint})}{collateral_B}\leq CR_s
    \end{align}
    \begin{align}\label{eq:2-5}
    \epsilon=1+\frac{donate}{supply_{IB}-borrowable_{IB}}
    \end{align}
Simultaneously, the collateral of Contract A is:
\vspace{-0.15cm}
    \begin{align}\label{eq:2-6}
    \epsilon\cdot[init_{mint}+(iter - 1)\cdot(borrowable_{IB}+init_{mint})]
    \end{align}
The final profit stage includes the redemption of withdrawable $token_{IB}$ and exhausting the lending market by borrowing all other assets. The profit will be as shown in Eq.~\ref{eq:2-7}. Setting Eq.~\ref{eq:2-7} equal to $borrowable_{s}+collateral_B$ and enabling Eq.~\ref{eq:2-5} hold equility, we can solve for $\epsilon$ as in Eq.~\ref{eq:2-8}.
    \begin{align}\label{eq:2-7}
    CR_{IB}\cdot\epsilon\cdot[iter\cdot init_{mint}+(iter-1)\cdot borrowable_{IB})]\\
    \label{eq:2-8}
    \epsilon=\frac{borrowable_{s}+\frac{iter\cdot (borrowable_{IB}+init_{mint})}{CR_s}}{CR_{IB}\cdot [iter\cdot init_{mint}+(iter-1)\cdot borrowable_{IB}]}
    \end{align}
The final profit for the attacker is:
\vspace{-0.15cm}
    \begin{align}\label{eq:2-9}
        &borrowable_{s}+collateral_A+borrowable_{IB}+\notag\\
        &init_{mint}-flash_{total}/(1-flash_{fee})\\
    \text{where}\ \  &flash_{total} =init_{mint} + collateral_A + donate\notag\\
    \text{and} \ \ &donate=(\epsilon-1)(supply_{IB}-borrowable_{IB})\notag
    \end{align}
\vspace{-0.55cm}
\section{Enhancement of The Attacks}\label{sec:enhance}
\subsection{Enhancement of the \textit{B\&B Attack}}\label{en:bb}
In order to escalate the profit potential of the \textit{B\&B Attack}, the hacker, acting as a secondary \textit{trader}, could execute reversed operations following the completion of the initial attack sequence, as shown in Figure \ref{fig:BB}. At this juncture, the lending market is saturated with a substantial volume of manipulated assets. Taking these manipulated assets as the unit of valuation, the price of stable assets can be effectively adjusted using similar attack methodologies. The profits of the reversed operations originate from the overestimated price for token A after the initial attack. Thus, in this reversed round of attack, we can perceive the stable assets as the \textit{manipulated assets}.

If we designate the initial attack as \textit{Attack Round -  \uppercase\expandafter{\romannumeral1}} and the subsequent reversed operation as \textit{Attack Round -  \uppercase\expandafter{\romannumeral2}}, the hacker has the potential to amplify their profits by cycling through numerous attack rounds. Here, even rounds enact forward operations, and odd rounds undertake backward operations.

\textbf{Mathematical Analysis:}
When considering iterated attack rounds, we can reflect on the transformation of state from the initial to the final state. In an ideal scenario, all manipulated assets in the lending market are borrowed and sold after an even number of rounds, leaving only stable assets in the market. When conducting a new round becomes unprofitable (Eq.~\ref{eq:1-5} in section \ref{sec:history} cannot be concurrently satisfied), we deduce:
\vspace{-0.15cm}
    \begin{align}\label{eq:3-1}
    \text{stable assets left}\le(CR_m-3+\frac{2}{CR_m})\cdot \frac{L_0}{1+\frac{init_m}{L_0}}
    \end{align}
To maximize the attacker's profit, the hacker could implement several rounds until Eq.~\ref{eq:3-1} holds true. The hacker's maximized profit is given by equation Eq.~\ref{eq:3-2}:
\vspace{-0.15cm}
    \begin{align}
    &init_s+\frac{init_m}{1+\frac{ init_m}{L_0}}-\text{stable assets left}= \notag \\ 
    \label{eq:3-2}
    &init_s+\frac{init_m}{1+\frac{init_m}{L_0}}-(CR_m-3+\frac{2}{CR_m})\cdot \frac{L_0}{ 1+\frac{init_m}{L_0}}
    \end{align}
In practical situations, due to the discrete nature of the number of rounds, the final result may be slightly different but still significantly improved compared to the pre-modification results.

\vspace{-0.2cm}
 \subsection{Enhancement of the \textit{B\&D} Attack}\label{subsec:bdEnc}
\begin{center}
    \begin{figure}[b]
  \raggedright
  \centering
  \includegraphics[width=0.45 \textwidth]{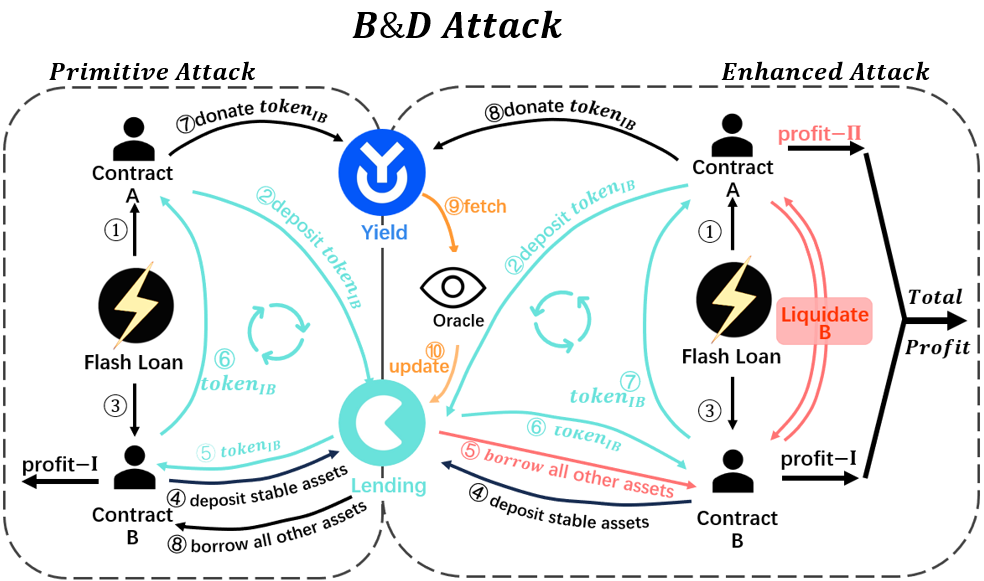} 
  \vspace{-0.35cm}
  \caption{The attack process of a \textit{B\&D Attack}. The primitive attack is analyzed in Section \ref{subsec:bd} while the enhanced attack is analyzed in Section \ref{subsec:bdEnc}.}
  \vspace{-0.15cm}
  \label{fig:BD}
\end{figure}
\end{center}

In this section, we will explain two key operations targeted at increasing the hacker's earnings by altering operation sequences and leveraging the liquidation process, as illustrated in Figure \ref{fig:BD}.

\textbf{Adjustment of Operation Sequences}
Profits can be increased by draining the lending market before the donation step, directly after the iterated mint and borrow procedure (process \ref{att:2-B} in Section \ref{subsec:bd}). Applying this adjustment, contract A's debt (Eq.~\ref{eq:2-2} in \ref{subsec:bd}) becomes Eq.~\ref{eq:4-1}, and its collateral before donation becomes Eq.~\ref{eq:4-2}.
\vspace{-0.15cm}
    \begin{align}\label{eq:4-1}
    &borrowable_s + iter\cdot (borrowable_{IB}+init_{mint})\\
    \label{eq:4-2}
    &\frac{iter\cdot (borrowable_{IB}+init_{mint})+borrowable_s}{collateral_B}\leq CR_{IB}
    \end{align}

\vspace{-0.15cm}
\textbf{Liquidation}
Profit can also be increased by adding a new \textit{liquidator} role. Contract A can liquidate contract B when the attack is completed, receiving the collateral assets of contract B with a liquidation incentive. The money required to liquidate contract B and contract A's collateral, in addition to the remaining liquidation funds, will be equivalent, as shown in Eq.~\ref{eq:4-4}. For our purposes, we'll call the liquidation incentive component $liq_{incentive}$.
\begin{align}\label{eq:4-4}
    &collateral_B\cdot (1-liq_{incentive})\notag \\
    =&\epsilon\cdot [init_{mint}+(iter-1)\cdot (borrowable_{IB}+init_{mint})]
\end{align}
With these enhancements, the final maximized profit for the hacker is represented by Eq.~\ref{eq:4-5}.
\vspace{-0.15cm}
{%
\allowdisplaybreaks
\begin{align}\label{eq:4-5}
    &borrowable_s+collateral_B+(borrowable_{IB}+init_{mint})\notag \\&-flash_{total}/(1-flash_{fee}) \\
    \text{where } &flash_{total}=init_{mint}+collateral_B+donate\notag \\
    \text{and } &collateral_B=\frac{iter\cdot (borrowable_{IB}+init_{mint})+borrowable_s}{CR_s}\notag \\
    \text{and } &\epsilon=\frac{collateral_B\cdot (1-liq_{incentive})}{init_{mint}+(iter-1)\cdot (borrowable_{IB}+init_{mint})}\notag \\
    \text{and } & donate=(\epsilon-1)\cdot (supply_{IB}-borrowable_{IB})\notag
\end{align}
}
We can find Eq.~\ref{eq:4-5} can be maximized by setting $init_{mint}=0$ because we can increase $iter$ to reduce flashloan fees. However, due to the gas limit, we can take a proper $init_{mint}$ to trade-off with $iter$.
\vspace{-0.4cm}
\subsection{Implementation}\label{subsec:imple}
With the toolchain \textit{Foundry}~\cite{Foundry}, we reconstructed the Agora Lending exploit and implemented the improved attack, which resulted in a profit of \$31.34M and \$35.10M, respectively, illustrating the enhancement of \$3.76M (12.0\%) with our refinement. We also improved the attack of the Loadstar to lower the flashloan fees and donation amount, improving the hacker's profit from around \$6.5M to \$9.84M by \$3.34M (51.4\%).
\vspace{-0.2cm}
\section{Discussion}
\label{sec:discussion}
\textbf{Post-Attack Value Extraction.} Our research, along with others~\cite{werner2022sok, qin2021attacking, Zhou2021OnTJ}, has highlighted a pattern where attackers frequently don’t exploit all their available opportunities, a concept known as `leaving money on the table'. This residual value can be a potential target for subsequent attackers who copy the initial exploit. Intriguingly, this scenario also presents an opportunity for the compromised project to deploy emergency measures, such as pausing operations or activating an emergency exit. A promising direction for future research is the in-depth analysis of these potential tactics, especially the potential for rescuing funds via back-running mechanisms~\cite{cryptoeprint:2023/892}.


\textbf{Mitigation Methods.} We propose future research in the following areas to strengthen defenses against Role-Play Attacks:
\vspace{-0.1cm}
\begin{itemize}[leftmargin=*]
    \item \textbf{Tailored Vulnerability Detection Toolkits:} Improve the security tools, like formal verification and fuzzing, used to identify Role-Play attacks. These tools can help auditors and project developers find and fix vulnerabilities before an attack occurs.
    \item \textbf{Real-Time Anomaly Detection Systems:} Explore the creation of real-time detection tools~\cite{291271, qin2023auto, gai2023blockchain} capable of identifying unusual role actions. These could signal an ongoing attack, possibly leveraging heuristics or machine learning algorithms.
    \item \textbf{Design Modifications:} Research into DeFi protocol design choices that limit a single transaction from playing multiple roles can be pivotal. While this may cause inconvenience, it can reduce the risk of Role-Play Attacks at the design level.
\end{itemize}

\vspace{-0.15cm}
\section{Conclusion}\label{sec:conclusion}
This paper delves into a recurrent attack strategy we define as the \textit{Role-Play Attack}.
We first propose a formal definition and categorize this attack based on the scenarios and \textit{roles}.
We analyzed 14 attack incidents that resulted in \$435.1M in losses, demonstrating the huge threat of this attack.
Furthermore, we offer a comprehensive mathematical analysis of two distinct attack methods of the \textit{Role-Play Attack}: the \textit{Borrow-And-Buy} Attack and the \textit{Borrow-And-Donate} Attack. Our research proposes enhancements to these specific attacks, potentially increasing their profitability by \textbf{\$3.34M (51.4\%)} and \textbf{\$3.76M (12.0\%)}, respectively.

\bibliographystyle{unsrt}
\balance
\bibliography{references}
\appendix

\end{document}